\documentclass[journal]{IEEEtran}
\usepackage{xcolor}

\ifCLASSINFOpdf
   \usepackage[pdftex]{graphicx}
\else
\fi
\usepackage{amsmath}
\usepackage{amssymb}

\begin{document}
%

%
%
\title{Explainable and Position-Aware Learning in Digital Pathology}
%
\author{Milan Aryal and Nasim Yahyasoltani 
\thanks{ The authors are with the Department
of Computer Science,  Marquette University, Milwaukee, WI, USA (e-mail: \{Milan.aryal, nasim.yahyasoltani\}@marquette.edu).}
\thanks{Part of this work was presented at the IEEE ICASSP Conference held in Greece, during June 4-10, 2023.}}

\maketitle
\begin{abstract}
Encoding whole slide images (WSI) as graphs is well motivated since it makes it possible for the gigapixel resolution WSI to be represented in its entirety for the purpose of graph learning.  To this end, WSIs can be broken into smaller patches that represent the nodes of the graph. Then, graph-based learning methods can be utilized for the grading and classification of cancer. Message passing among neighboring nodes is the foundation of graph-based learning methods.  However, they do not take into consideration any positional information for any of the patches, and if two patches are found in topologically isomorphic neighborhoods, their embeddings are nearly similar to one another. 
 In this work, classification of cancer from WSIs is performed with positional embedding and graph attention.  In order to represent the positional embedding of the nodes in graph classification, the proposed method makes use of spline convolutional neural networks (CNN).  The algorithm is then tested with the WSI dataset for grading prostate cancer and kidney cancer. A comparison of the proposed method with leading approaches in cancer diagnosis and grading verify improved performance. The identification of cancerous regions in WSIs is another critical task in cancer diagnosis. In this work, the explainability of the proposed model is also addressed. A gradient-based explainbility approach is used to generate the saliency mapping for the WSIs. This can be used to look into regions of WSI that are responsible for cancer diagnosis thus rendering the proposed model explainable.
\end{abstract}
\section{Introduction}
\label{sec:intro}
To diagnose and classify cancer, a tissue sample or excision need to be evaluated by a pathologist.  To this end, tissue samples are processed and cut into small layers before being put on a glass slide and are then stained using histological hematoxylin and eosin (H\&E). The pathologist can diagnose any malignancy by scanning  and examining the slides under the microscope. Digitized versions of H\&E slides are known as WSIs~\cite{Aeffner2019-vo} where they can be digitized at different microscopic resolutions.

 Despite the high-resolution and computational complexity of WSIs, their use in computational pathology and machine-based cancer diagnosis is increasing. In fact, recent advances in deep learning has facilitated the adoption of WSIs for the computer-aided diagnosis and grading of cancers. However, even with recent progresses made in computational methods for digital pathology, processing of WSIs for training in deep learning is considered a challenging task. The fact that WSIs have billions of pixels in a single file typically larger than a gigabyte, makes it challenging to train with common methods such as CNN.
 
Recently, multi-instance learning (MIL) approaches have been shown to be very successful in learning of the WSIs~\cite{multiinstance},~\cite{MIL1},~\cite{mil2},~\cite{nature-weaklyMIL},~\cite{Lu19}. In MIL, instead of learning individual labels, the instances are bagged together and the training is performed on them. More specifically,  the WSI can be broken into many patches and then bagged. In the case of binary classification tasks such as cancer diagnosis, if the bag consists of a single patch of cancerous tissue, the WSI is labeled as positive; otherwise it is categorized as negative. Once the patches are bagged together they are passed through CNN and transfer learning for training. Although, the accuracy obtained through this method is very promising, there are limitations that need to be considered. For example, MIL does not take into account the neighboring patches and some patches in WSI may not get selected for training. 

Currently, the research in WSI learning is shifting from MIL learning to graph-based learning~\cite{ma},~\cite{10.1117/12.2550114},~\cite{chen2021whole},~\cite{Ding2020FeatureEnhancedGN}. Representation learning in graphs is performed through an encoder that maps nodes to a low-dimensional embedding space. Encoders encode nodes as low-dimensional vectors that summarize their local neighborhood. These embeddings enable tasks like node classification, link prediction, graph classification, and graph generation~\cite{NIPS2017_5dd9db5e},~\cite{kipf2017semi},~\cite{NEURIPS2018_e77dbaf6}. Recently, graph neural networks (GNNs) has become a very popular way to encode nodes for embedding. To learn node embeddings, GNNs take into account the local neighborhood information of a node.
The advantage of graph-based learning over MIL is that full-resolution WSI can be represented as a graph, thus encoding tumor neighborhood information. 
Then, the whole graph is learnt through a variant of message passing algorithms followed by the global aggregation method and multi-layer perceptron (MLP) for the cancer classification.

\begin{figure*}[htp]
    \centering
    \includegraphics[width=\textwidth]{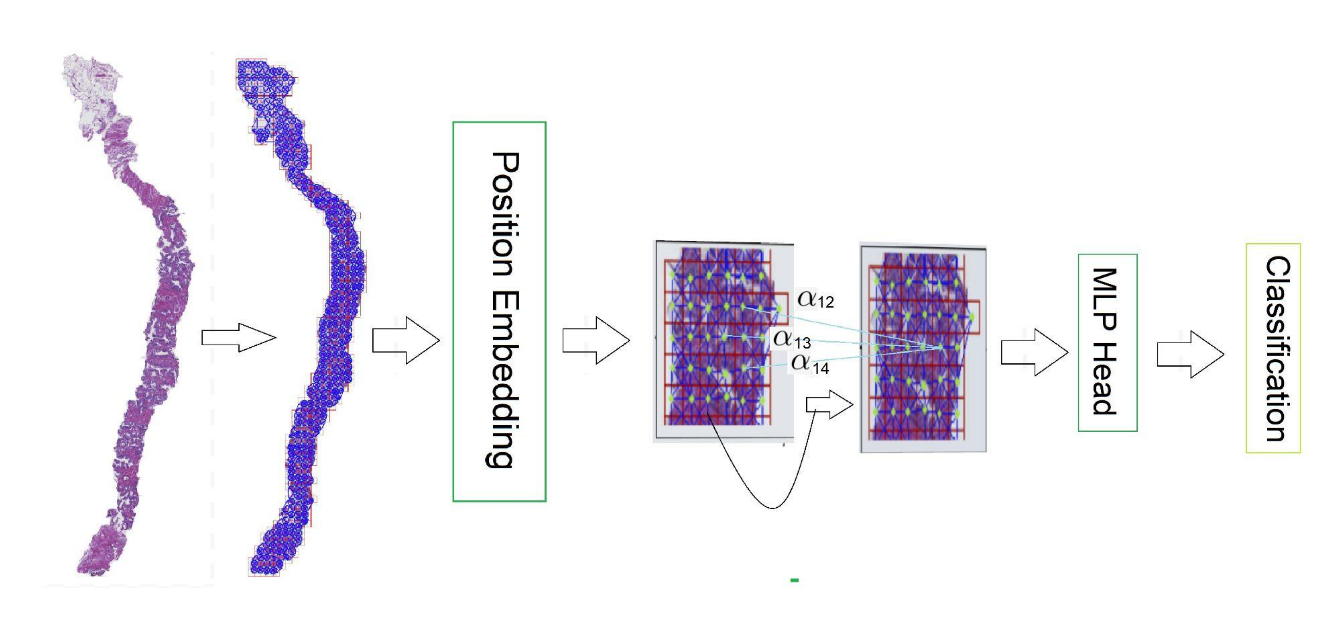}
    \caption{The proposed model's system block diagram. A graph is constructed from the input WSI. Then, the position embedding is learnt for the input graph. The graph is then passed through 2 layers of graph attention. The attention layer is followed by MLP for the classification.}
    \label{fig:diag}
\end{figure*} 

Despite all the recent advances in the adoption of GNN in WSI classification, they fall short in distinguishing nodes with similar local neighborhoods {~\cite{ma},~\cite{10.1117/12.2550114},~\cite{chen2021whole},~\cite{Ding2020FeatureEnhancedGN},~\cite{9156339},~\cite{801842acbabe476990854c32fdfdbde5},~\cite{valentin}}. In fact, the position of the nodes with respect to the graph is not captured in traditional GNN and if two nodes are located in topologically isomorphic neighborhoods, their embedding is almost identical. To this end, modifying GNN such that it takes into account node features and captures the position of the nodes is well motivated for improving representation learning~\cite {position}. 


With WSI broken into patches, one needs to take into account the spatial information of each patch and capture the information between those patches. This can be done by capturing the position embedding of each patch. Through an spline CNN algorithm~\cite{splinecnn}, we aim at capturing the position embedding of patches. Once the position embedding of the patches is captured, we then use an attention-based method to perform the message passing among patches. As not all patches are cancerous in WSI, we want to learn differently from each patch and give different attention to it. This way, we can identify cancerous regions, more efficiently.

The degree to which the deep learning model can be explained is of critical significance. The more readily explicable a model is, the more effectively it may be interpreted. When the model's interpretability and explainability are improved, the prediction that the model generates can be comprehended more easily. This helps in constructing models that are more trustworthy, and thus more applicable in medical diagnosis. Simple machine learning models, such as linear regression, are highly interpretable, but deep learning models are generally less interpretable due to their complexity.  However, deeper models often offer a higher accuracy compared to simpler models~\cite{gcam}. Recently, there has been more focus on developing algorithms to make complex models explainable.  The use of CNN has huge success in vision related task such as image classification, segmentation and object detection. The explainablity of CNN-based algorithms has been widely explored in recent years. In~\cite{cam},~\cite{gcam}, ~\cite{cam2},~\cite{cam3} CNN model is made explainbale by visualizing the area of interest in image classification. The explainability model for CNN model has been expanded to graph-based models as well in~\cite{pope_explain}. In graph-based learning the explainabilty amounts to identifying the importance of each node in prediction. 

The explainability of a model is of paramount importance in the context of cancer detection, as it helps to increase trust in the model's predictions. When working with WSI, it is essential to pinpoint the particular areas of the image that are accountable for making the cancer diagnosis. Because malignant cells may be concentrated in some areas, these regions might not necessarily be present throughout the entirety of the image. In this work, the goal is to use an explainability method that is based on graph neural networks in order to locate areas related top malignancy inside WSIs. By identifying the specific regions of the image that are responsible for the cancer diagnosis, the model can provide more accurate and trustworthy predictions.

To summarize, the contribution of this work is three-fold: 1) It introduces a graph-based self-supervised position-aware algorithm for cancer diagnosis using WSIs; 2) To encode position information of the patches in the WSIs obtained through spline CNN, an attention network is proposed to promote different weights on cancerous versus non-cancerous nodes; and 3) Based on gradient-weighted class activation mapping (Grad-CAM), an explainability algorithm for the proposed model is developed. Using Grad-CAM the heatmaps are generated to identify the regions of interest in WSIs which can be later used to visualize and detect the cancerous regions.


This paper is organized as follows. Section~\ref{sec::rw} addresses the related work. In section \ref{sec::Method}, the proposed method is presented. The implementation details are shown in section~\ref{sec:implementation}. Section~\ref{sec:results} presents the performance of the algorithm for different types of cancer WSIs and discussion of the results is followed in section~\ref{dis}. The conclusion is provided in section~\ref{sec:conc}.

\begin{figure*}[htp]
    \centering
    \includegraphics[width=\linewidth]{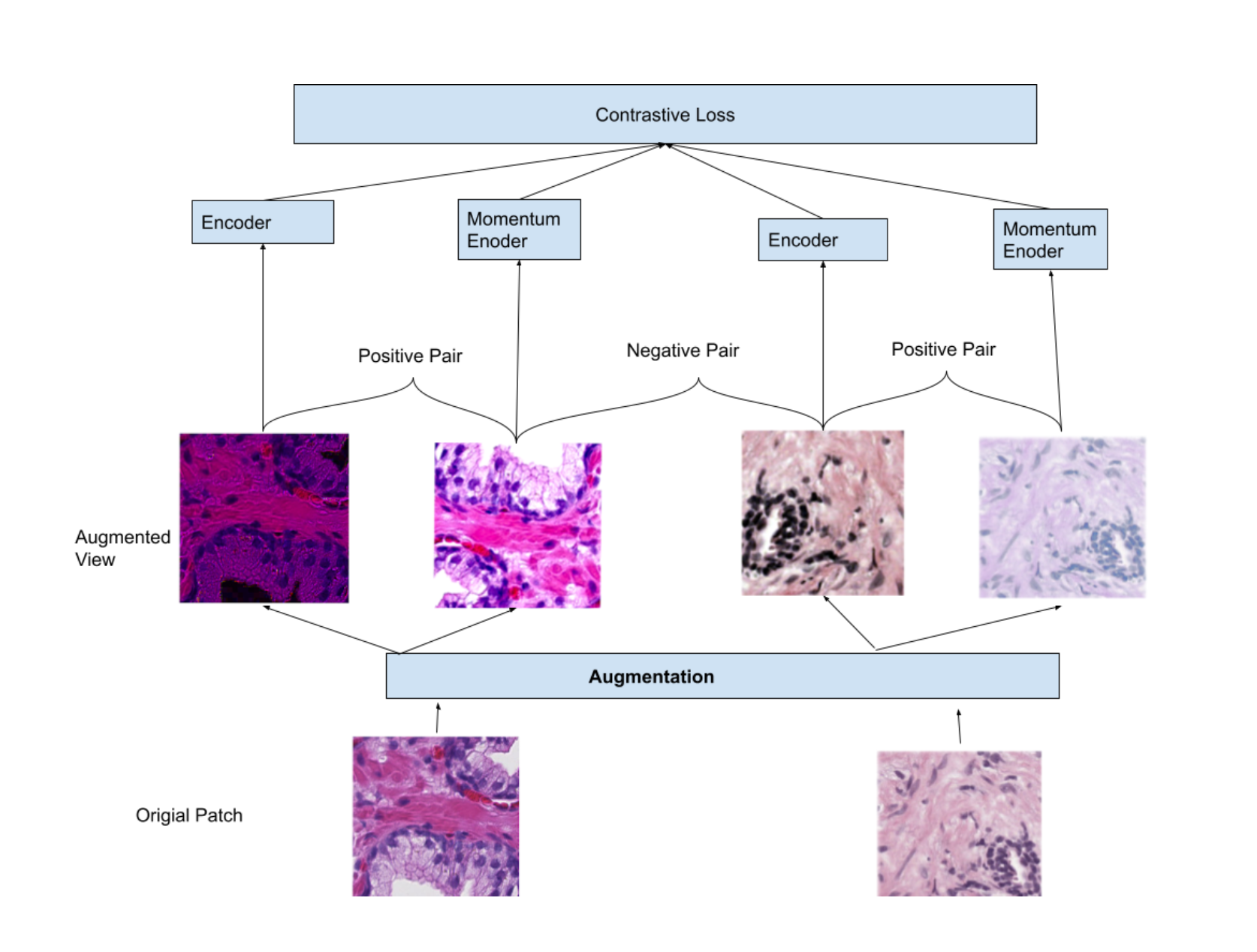}
    \caption{Contrastive learning approach for extracting features in patches.}
    \label{fig:ssl}
\end{figure*}

\section{Related Work}
\label{sec::rw}
The use of GNN-based methods are increasing in WSI for cancer diagnosis. The WSIs are images of digitized histopathological slides and are not originally represented as graphs. So, the first challenge in using graph-based learning approaches is to find a way to represent the image data as a graph. Once WSIs are encoded as graphs, different graph representation learning methods can be deployed. The common approach to construct the graph is to form cell graph ~\cite{graphm}. {In this method, first the nuclei in the WSIs are identified and represented as the nodes in the graph. Then, the graph is formed with edges connecting other neighboring nuclei in the tissue.} Another approach is to form patch-graph method. In this method, the WSI is broken into patches as nodes and the graph is constructed from those nodes.  

In~\cite{10.1117/12.2550114}, the authors addressed the classification of WSIs for breast cancer using graph convoutional network (GCN)~\cite{kipf2017semi}. More specifically, the WSIs were constructed as graphs using the constituents cells with nuclear morphology as vertex features and the gland formation as edge attributes. In~\cite{801842acbabe476990854c32fdfdbde5}, a model for gleason score grading of prostate cancer has been proposed. The authors constructed a graph with nuclei as the nodes and the distance between each node as the edge. The features of each node are extracted using contrastive predictive coding and morphological features. Then, this graph was passed into GCN for the grading of high-risk and low-risk cancer patients. These methods use cell nuclei for the construction of the graphs from WSIs.

Other methods used in the graph construction from WSIs include patch-based methods.  In~\cite{9156339} GCN-based multi-instance learning method is implemented for the classification of lymph node. First, the patches with regions of interest are selected and bagged together. Then, this bag of patches is passed into auto-encoder and GCN for feature selection and classification, respectively.
In~\cite{Ding2020FeatureEnhancedGN} the genetic mutation of colon cancer is modeled. Randomly selected patches from each WSIs act as nodes and CNN is used to extract features for each node. The constructed graph is passed through spatial-GCN to predict the mutations.  Survival analysis is performed using graph-based methods in~\cite{chen2021whole}.

The existing methods for the graph-based learning of WSIs use popular message passing techniques such as GCN and {graph attention network (GAT)}, which do not take node position in the graph into account. In this work, position embedding is also captured. 

Explainability is another aspect in improving the interpretability of a model. This makes it easy to explain the model and its outcome. The literature in explainabilty of graph-based models in WSIs is very limited. In~\cite{graphex} the authors have presented an explainability method based on the importance of the nuclei for cancer classification.

\section{Method}
\label{sec::Method}
The system diagram of the proposed model is presented in Fig.~\ref{fig:diag}. The process begins by converting the WSI into a graph, with each patch of the WSI being represented as a node and 
edges connecting those nodes. In this work, the model aims to learn the position of each node in the graph. This allows to maintain the spatial information of each patch in WSI even though it is represented as graph data. We deploy Spline CNN to learn the position of each node in the WSI ~\cite{splinecnn}.

After capturing the position embedding, the graph is passed through the attention network. The model consists of 2 layers of attention network. The first layer of attention network consists of multi-headed attention followed by single attention head network. Following the attention layer, the model classifies the WSI using the graph classification. The results from attention layers are global pooled using mean pooling. This layer is forwarded to multi-layer perceptron network and then to a classification layer for cancer grading.

\subsection{Graph construction and feature extraction}
The graph data structure is represented as $G=(V,E,X)$ where $V$, $E$ and $X$ denote the set of nodes, the edge set, and node features, respectively. It is necessary to transform the WSI into graphs in order for each WSI to be represented as $G=(V,E,X)$.

The aim of using graph-based learning for WSIs is to learn from the whole resolution of WSI. So, the graph is constructed from the highest level of resolution of WSI. The graph formation of WSI is implemented by breaking it into smaller patches. In this work, WSI is broken into non-overlapping patches of size $256\times256$. Since each patch acts as a node in the graph, its position in the WSI is used to construct connectivity between corresponding neighbors. Based on fast approximate k-nearest neighbor~\cite{Muja09fastapproximate}, the connectivity between nodes are established. Since each WSI is different in size, the number of nodes in each graph is different. For the constructed graph in this work, we consider $V$ as the set of patches (nodes) and the edge set $E$ is given by connection between those patches. 
With $N$ as the number of nodes in the graph $G$, the set of features for each node, i.e., $X\triangleq\{\boldsymbol{x}_v\}_{v=1}^{N}$ is obtained by extracting features from each node, where $\boldsymbol{x}_v$ is the feature vector of node $v$. For feature extraction of each patch, self-supervised learning method is deployed. Typically, medical data suffers from limited number of annotated data. The lack of annotated data makes self-supervised algorithms suitable for extracting features from patches. More specifically, contrastive learning framework is used to train patches~\cite{chen2020simple},~\cite{ma}. This way, feature extraction of patches is performed with an unsupervised learning approach.
In Fig.~\ref{fig:ssl}, the feature extraction method for patches based on contrastive learning is visualized. The contrastive learning approach maximizes the similarity between the positive pair and
dissimilarity among the negative pair. The augmentation is used to generate multiple view of the same images to create the positive pairs. When the network trains the augmented view of different patches, those are considered to be negative pairs.

\subsection{Position embedding in the graph}
Graph learning entails embedding nodes into lower dimensions. The node embedding is performed by an encoder based on different information about the node and its neighbors. The embedding space is optimized so that the distance between nodes captures the relative position of the original graph. The encoder function $f: V\xrightarrow{} Z$ maps nodes $v \in {V}$ to embeddings $Z\triangleq\{\boldsymbol{z}_v\}_{v=1}^{N} \in \mathbb{R}^d$, where $d$ refers to feature size in the embedding space.
The most common approaches for embedding include matrix factorization~\cite{fact1}, \cite{matfac1}, random walk and neural learning methods such as graph neural learning. In matrix factorization, node embedding is constructed based on node similarity matrix. Some of the popular random walk methods include DeepWalk~\cite{deepwalk}, node2vec~\cite{node2vec}. With capability to learn from node features and inductive learning, GNNs are most commonly used in node embedding. 
\begin{figure}[htp]
    \centering
    \includegraphics[width=\linewidth]{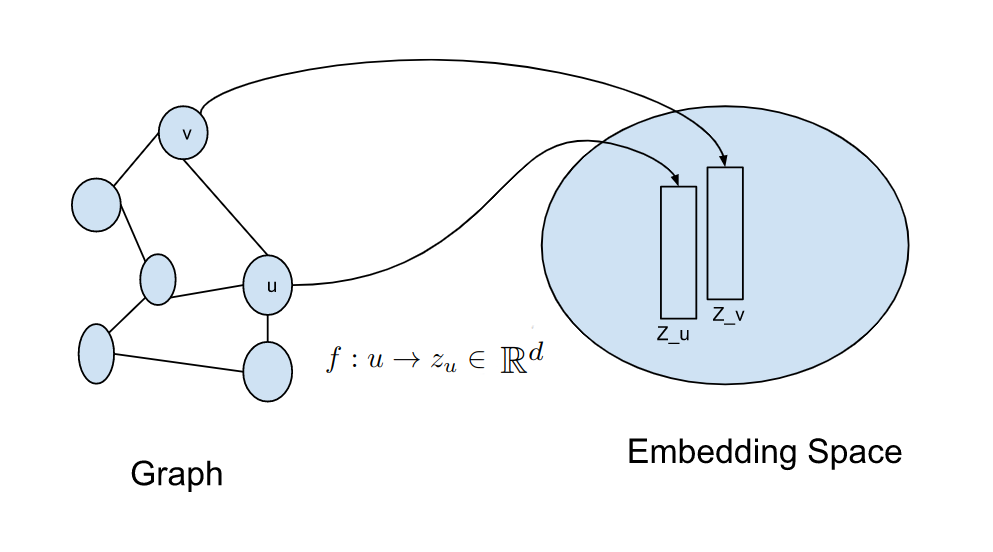}
    \caption{Embedding graph nodes into an embedding space.}
    \label{fig:embed}
\end{figure} 

Due to unstructured nature of graph data, the positions of the nodes are not naturally encoded in them. In~\cite{position}, it is shown that popular graph-based learning methods such as GNN are capable of capturing structural information, but the position of the node is not taken into account. As a result, the nodes with different labels having similar neighborhoods will end up with the same embedding and similar representations. To avoid this issue and make the learning aware of the position of the nodes in the graph, it is useful to have positional embedding. 

In this work, WSIs are represented with 2D coordinates and the goal is to classify each WSI incorporating the position of each patch in the WSI graph. The coordinates of each patch is the midpoint of each patch in the WSI. The proposed algorithm to capture the position of the nodes is based on spline CNN~\cite{spline},~\cite{spline2}.  Spline CNN uses x,y coordinates to learn the geometry of the graph. We deploy B-splines~\cite{piegl1996nurbs} as presented in~\cite{splinecnn} to learn nodes positions. In Spline CNN, with set of neighboring nodes defined as $N_i$, the feature vector of node $i$, denoted by $f_i$ is given by~\cite{spline},~\cite{splinecnn}:
\begin{equation}
\begin{aligned}
   f_i = \frac{1}{|N_i|} \sum_{j\in N_i} f_j\cdot g(u(i,j))
\end{aligned}
\end{equation}
where $g(u(i,j))$ is a kernel function based on the coordinates of nodes $i$ and $j$. Suppose the coordinate of node $i$ is given by $(x_i,y_i)$ and the coordinate of node $j$ is $(x_j,y_j)$, then $u(i,j) \triangleq (|x_j-x_i|,|y_j-y_i|) \triangleq (x,y)$. The kernel function is defined as follows:
\begin{equation}
\begin{aligned}
   g(u(i,j)) = \sum_{p\in \mathcal{P}} w_p \cdot N_{hori,p}(x) \cdot  N_{vert,p}(y)
\end{aligned}
\end{equation}
In the above equation, $N_{vert}$ and $N_{hori}$ are B-splines bases, $w_p$ is a trainable parameter for each element $p$ in the Cartesian product $\mathcal{P}$ of $N_{vert}$ and $N_{hori}$.

\subsection{Graph attention network}
As defined in section~\ref{sec::Method}, a graph can be represented as $G=(V,E,X)$ where $V$ denotes the set of nodes with number of nodes $N = |V|$ , $E$ refers to the set of edges. Then,  $X \triangleq \{\boldsymbol{x}_1,\boldsymbol{x}_2,\hdots,\boldsymbol{x}_N\} \in \mathbb{R}^d$ accounts for the nodes feature vectors. Graph learning entails embedding graph nodes to a lower dimensional space. 

GNN is one of the most popular embedding methods used for graph representation learning where the learning is based on passing and updating messages among the nodes, iteratively.

Let us denote the feature vector after message passing as $X^{'} = \{\boldsymbol{x}^{'}_{1},\boldsymbol{x}^{'}_2,\hdots,\boldsymbol{x}^{'}_{N}\}$ in the next layer. The message passing for each node in GNN~\cite{William} is based on the neighboring nodes and given by:

\begin{equation}
\begin{aligned}
    \boldsymbol{x}_i^{'} &= \sigma\left(W_{self}\boldsymbol{x}_i+W_{neigh}\sum_{j\in \mathcal{N}_i}\boldsymbol{x}_j+b\right) 
\end{aligned}
\end{equation}

where $\boldsymbol{x}_i$ is the feature vector of the node whose message is being updated, and $\boldsymbol{x}_j$ are the neighboring nodes feature vectors.  Then, $W_{self}, W_{neigh}$ are the trainable weight parameters updated during the message passing from itself and neighboring nodes, respectively and $\sigma$ is the activation function.

Graph convolutional neural network (GCN)~\cite{kipf2017semi}, as a variant of GNN is very similar to CNN. In GCN message passing aggregates the message from its neighboring nodes features.  The message update for the GCN follows:
\begin{equation}
\label{eq:GNN}
    \boldsymbol{x}_i^{'} = \sigma\left(\sum_{j\in \mathcal{N}_i}\frac{W\boldsymbol{x}_j}{\sqrt{|\mathcal{N}(j)||\mathcal{N}(i)}} \right)
\end{equation}
where $\sigma$ is the activation function and accounts for the non-linearity, and $W$ are the trainable parameters updated during the training of GCN. is The degree of neighboring node $j$ is denoted by $|\mathcal{N}(j)|$ and $|\mathcal{N}(i)|$ accounts for the degree of the node $i$ for which message passing is being performed.
 
Using GCN, the message passing to a node is achieved via its neighboring nodes and all the neighboring nodes are equally accountable. The graph attention network (GAT) modifies this convolutional operator by assigning attention weight in message passing stage among nodes. The attention mechanism allows the network to learn the specific weights from the specific nodes. This allows to learn the importance of each neighboring nodes in message passing.  

More specifically, the message passing in GAT~\cite{GAT} is given by:
\begin{equation}
\label{eq:gat}
    \boldsymbol{x}_i^{'} = \sigma\left(\sum_{j\in \mathcal{N}_i} \alpha _{ij}W\boldsymbol{x}_j\right)
\end{equation}
where $\alpha _{ij}$ is the attention weight between nodes $j$ and node $i$. The attention weights are given by:

\begin{equation}
\label{eq:att}
    \alpha_{ij} = \frac{exp(LeakyReLU(a^T[W\boldsymbol{x}_i||W\boldsymbol{x}_j]))}{\sum_{k\in\mathcal{N}}exp(LeakyReLU(a^T[W\boldsymbol{x}_i||W\boldsymbol{x}_j]))}
\end{equation}

where $||$ is the concatenation operator and Leaky Rectified Linear Unit (LeakyReLU) was used as activation function.  

The stabilization of the self-attention mechanism is obtained by adding the multi-head attention. This operation is achieved by applying $K$ independent attention-based operators and then concatenating them.  Adopting multi-head attention network with $K$ heads,~\eqref{eq:gat} is replaced by:
\begin{equation}
\label{eq:gathead}
    \boldsymbol{x}_i^{'} = \sigma\left(\frac{1}{K}\sum_{k=1}^{K}\sum_{j\in \mathcal{N}_i} \alpha _{ij}W^{k}\boldsymbol{x}_j\right)
\end{equation}
In this work, two layers of multi-head GAT is considered.

Use of attention-based method allows for different weight to nodes in the network.  In cancer diagnosis, through a graph-based WSI, neighboring nodes might consist of both cancerous and non-cancerous patches. To this end, assigning different importance weights to different nodes in a neighboring region in message passing is well motivated. 

 \subsection{Explainability of the model}
The application of GNN in cancer classification has the potential to be of significant assistance if the regions responsible for cancer classification can be pinpointed.  The physicians and healthcare team need a model to be interpretable, so that  the reasoning behind the suggested prediction can be well justified and understood. When it comes to the task of predicting cancer especially using deep learning models, explainability is of the paramount importance. When WSI is examined under the microscope by a pathologist, the pathologist is screening for possible areas inside the tissue that could be to blamed for cancer. The classification of WSI, together with the identification of the regions that causes a particular model for classification in WSI, could be of tremendous assistance to pathologists as well.

In this paper, we deploy explainability method implemented in graphs that are mainly tailored to CNN~\cite{gcam},~\cite{gcam1}. In particular,  graph Grad-CAM method for WSI is implemented. The choice of this approach is inspired by ~\cite{pope_explain} as the authors address explainability for molecular graphs.

As shown in ~\eqref{eq:GNN} in the layer $l$ of GNN, feature $k$ of the node $n$ is given by $\boldsymbol{x}_{k,n}^l$. Then, the average pooling of the feature at layer $l$ is given by:
\begin{equation}
    e_k = \frac{1}{N}\sum_{n=1}^N\boldsymbol{x}_{k,n}^l
\end{equation}
Based on this average pooling score $e_k$, and weights $w_k^c$ for predicting class $c$ from feature $k$, the class score $y_c$ can be calculated as shown below:
\begin{equation}
y_c = \sum_k w^c_k e_k
\end{equation}
The following equation for the Grad-CAM gives each feature importance score based on class and layer: 
\begin{equation}
    \alpha_{k}^{l,c}=\frac{1}{N}\sum_{n=1}^{N}\frac{\partial y^c}{\partial \boldsymbol{x}_{k,n}^l}
\end{equation}
Then, the heatmap for each node is generated by applying Rectified Linear Units (ReLU)~\cite{relu} functions as:
\begin{equation}
    L^c_{Grad-CAM}[l,n] = ReLU\left(\sum_k \alpha_{k}^{l,c} \boldsymbol{x}_{k,n}^l \right)
\end{equation}

The advantage of using Grad-CAM is that the heatmaps for the nodes can be calculated from any layer of the network. 
\begin{figure*}[btp]
    \centering
    \begin{tabular}{c}
          \includegraphics[width=\textwidth]{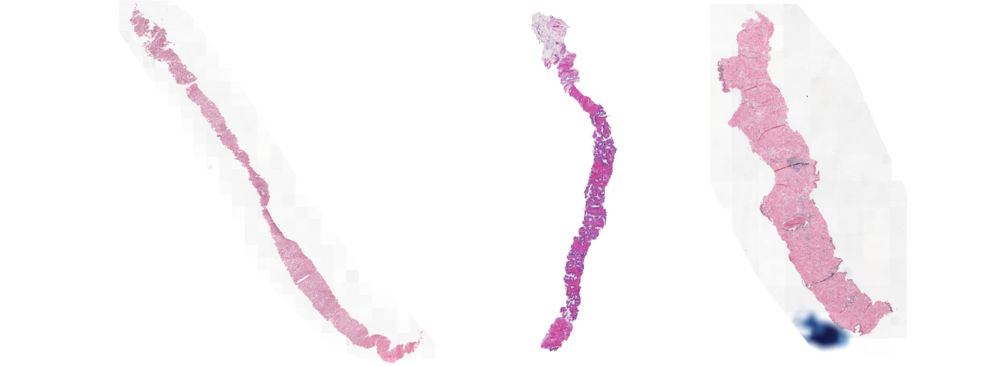} \\
         \small (a) Sample WSIs from Prostate Cancer Dataset \\
         \includegraphics[width=\textwidth]{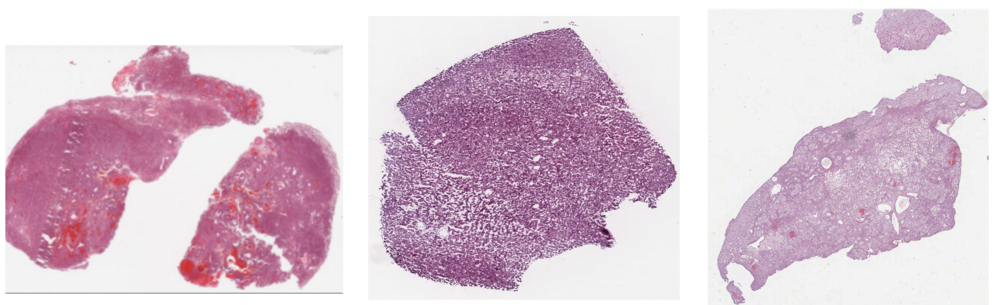} \\
         \small (a) Sample WSIs from Kidney Cancer Dataset
  \end{tabular}
  \medskip

  \caption{Sample WSIs from dataset used in simulation.}
  \label{WSI}
\end{figure*}

\section{Implementation Details}
\label{sec:implementation}
\subsection{Dataset}
Using two different datasets, we examined how well the proposed position-aware and graph attention-based model works for grading WSIs. To be more specific, we tested the performance of models with samples from both prostate and kidney cancers. After lung cancer, prostate cancer is the type of cancer that is responsible for the second highest number of deaths in men~\cite{cancer_stat}. Kidney cancer is a prevalent form of cancer that can impact both sex and ranks in the top 10 cancers that result in mortality~\cite{cancer_stat}. The prostate dataset consists of biopsies from the WSIs taken from biopsies from the prostate glands and the WSIs for kidney dataset are obtained from the kidney resections (larger samples). The sample WSIs of each dataset are shown in Fig.~\ref{WSI}. It can been seen that the structure of a biopsy is very different from a kidney resection as shown in Fig.~\ref{WSI}, 

 The Cancer Genome Atlas (TCGA) has been used for the kidney cancer dataset~\cite{TCGA}. The TCGA portal provides various kinds of data for cancer research, such as demographics of the patient, WSIs, and biomarkers.  The data related to kidney cancer consists of the most prominent type of kidney cancer, renal cell carcinoma, which is addressed in this paper. In kidney cancer, the most prevalent malignancy Renal Cell Carcinoma~\cite{Hsieh2017-mm} accounts for more than 90\% of cases. In our work, three major types of RCC  grouped as  KIRC (Kidney Renal Clear Cell Carcinoma), KIRP (Kidney Renal Papillay Cell Carcinoma) and KICH (Kidney Chromophobe) are considered. 
The WSI dataset related to prostate cancer has been accessed from ~\cite{kaggle1}. In prostate cancer, we look into the problem of grading the prostate cancer into different ISUP grades. ISUP grades consist of 5 grades of Gleason Grading (GG)~\cite{Chen2016-cl} for the prostate cancer.
In Table~\ref{tab:1} the number of samples of each gleason grading for prostate cancer is listed.  Similarly, Table~\ref{tab:2} shows the number samples of each type of cancer in kidney dataset.

\begin{table}[h]
\caption{Number of samples in prostate cancer dataset}
\begin{center}
\begin{tabular}{|c|c|}
\hline
\textbf{Description }& \textbf{Number of samples} \\
\hline
GG1 & 2666  \\
\hline
GG2 & 1343  \\
\hline
GG3 & 1250\\
\hline
GG4 & 1242  \\
\hline
GG5 & 1224\\
\hline
\end{tabular}
\label{tab:1}
\end{center}
\end{table}
\begin{table}[h]
\caption{Number of samples in kidney cancer dataset}
\begin{center}
\begin{tabular}{|c|c|}
\hline
\textbf{Description }& \textbf{Number of samples} \\
\hline
KIRC & 519\\
\hline
KIRP & 300 \\
\hline
KICH & 121 \\
\hline
\end{tabular}
\label{tab:2}
\end{center}
\end{table}
\begin{table*}[h]
\caption{Comparison of Kappa score for prostate cancer and kidney cancer datasets.}
\begin{center}
\begin{tabular}{|c|c|c|}
\hline
\textbf{Method}& \textbf{Prostate Dataset} & \textbf{Kidney Dataset} \\
\hline 
MIL with Efficinet Net~\cite{mil_2018} & 0.87 & 0.868 \\
\hline
GCN based model~\cite{ma} & 0.899 & 0.939 \\
\hline
Proposed model  & 0.912 & 0.941 \\
\hline
\end{tabular}
\label{tab:3}
\end{center}
\end{table*}

\subsection{Training settings}
Initially, the patches were trained for feature extraction for graph learning. To handle unlabeled patches, contrastive learning as a self-supervised method was used to train the patches. Then,  MOCOv3~\cite{mocov3} with ResNet50~\cite{resnet} as backbone was deployed to train the patches for feature extraction. The training was performed for 75 epochs with initial learning rate of $1\times10^{-3}$ and the weight decay of $1\times10^{-4}$. Cosine scheduler was chosen to update the learning rate during the training.

For graph learning training, the network consists of two layers of Spline CNN followed by two layers of graph attention. This is then forwarded to a MLP for cancer grading. The training was done for 20 epochs with initial learning rate of $1\times10^{-3}$ and weight decay of $5\times10^{-4}$. The AdamW~\cite{adamw} optimizer was used for training. The batch size for each epoch was chosen as 2. We use Kappa score~\cite{kappa} to measure the performance of the model. 
\begin{figure*}[htp]
    \begin{tabular}{p{8cm}p{30cm}p{18cm}}\\
    \multicolumn{3}{c}{\includegraphics[width=\textwidth]{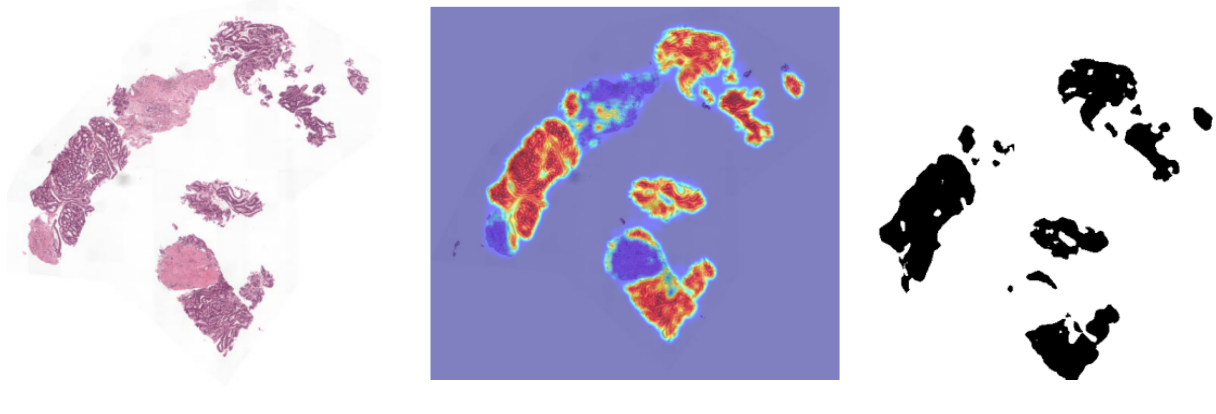}}\\
    \multicolumn{1}{c}{\centering(i)} & \multicolumn{1}{c}{\centering(ii)} & \multicolumn{1}{c}{\centering(iii)}\\
    \multicolumn{3}{c}{\small(a) Activation map generated for the cancer with ISUP grade 4}\\
      \multicolumn{3}{c}{\includegraphics[width=\textwidth]{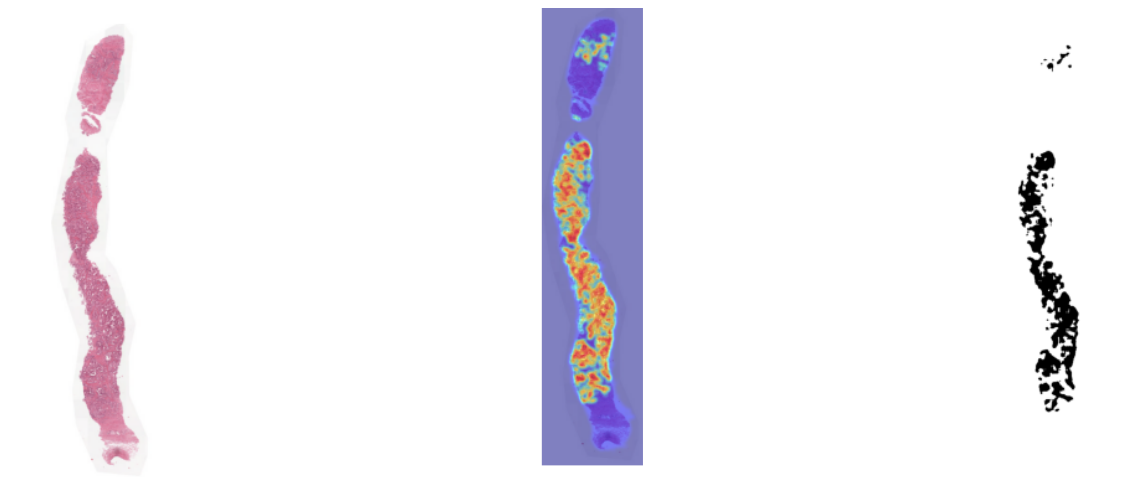}}\\
        \multicolumn{1}{c}{\centering(i)} & \multicolumn{1}{c}{\centering(ii)} & \multicolumn{1}{c}{\centering(iii)}\\
        \multicolumn{3}{c}{\small (b) Activation map generated for the cancer with ISUP grade 5}\\
\end{tabular}
  \medskip
    \caption{Explainability of the model. The heatmap is generated using the Grad-CAM. The activation map is compared with annotation provided for the WSI. (i) WSI (ii) Activation map generated by the model (iii) Annotation for the WSI. }
    \label{fig:Annotation}
\end{figure*} 

\section{Results}
\label{sec:results}
The proposed approach was evaluated using the two different cancer datasets. The performance of the proposed model is compared with self-supervised GCN-based model~\cite{ma} and MIL-based leading methods~\cite{mil_2018}. 

The position-aware method that was proposed for grading and classifying cancer was initially evaluated using prostate cancer dataset. To this end, a classification of cancer based on one of the the five ISUP grades can be obtained. On this dataset, the proposed approach yields a Kappa score of 0.912 as its outcome. Evaluating the cancer diagnosis and grading with a GCN-based and a MIL-based models for comparison, the Kappa scores are 0.899 and 0.87, respectively.

The proposed method was also applied to a dataset consisting of kidney cancer data.  The obtained kappa scores are very comparable. The proposed model obtaines a Kappa score of 0.941 when it was applied to the kidney dataset. In comparison, the Kappa score for the GCN-based and MIL-based models are 0.939, and 0.868, respectively. As can be seen in Table~\ref{tab:3}, the addition of patch position into WSI graph learning results in a significant improvement in cancer grading and classification.

In general, these results point to the fact that the proposed position-aware method is a promising approach for the diagnosis and treatment of cancer. This was verified with different types of WSI samples including both prostate biopsies and Kidney resections to classify tumors affecting the prostate and the kidney. 

\subsection{Explainability visualized to the WSI}
To better demonstrate the power of the proposed algorithm, it is crucial to investigate the explainabilty of the model. In other words, the expectation is that the algorithm can identify the regions and areas in the WSI that are responsible for the cancer diagnosis. This result verifies that the proposed model is interpretable. To this end, the heatmaps for each node is generated after the activation layer and is superimposed with the WSI. Then, the generated heatmap is compared with the original annotation provided in the dataset.

The heatmaps are generated for two cases of prostate cancer, one with an ISUP grade of 4 and the other with an ISUP grade of 5. In Fig.~\ref{fig:Annotation} the top image shows the heatmap for prostate cancer with an ISUP grade of 4, while the bottom image shows the heatmap for the prostate cancer with an ISUP grade of 4. The top left image (i) shows the  original WSI for ISUB grade 4. 
As we can see in the image the areas where the cancer is spread the tissue is not smooth and the growth pattern is irregular. This is further confirmed in the top right image in (iii) which shows the annotation provided for the WSI and clearly depicts the cancerous regions in the WSI. The image on the top middle (ii) is generated using Grad-CAM for ISUB grade 4 WSI and the red regions demonstrate the areas responsible for the cancer diagnosis. The identified regions by Grad-CAM are very aligned with the annotations provided by the pathologist as shown in (iii). 
Similar trends hold for WSI of ISUB grade 5, as shown in Fig 5, (b). As it can be seen, the annotation matches with the cancerous areas identified  by the Grad-CAM algorithm.

\section{Discussion}
\label{dis}



In this work, graph-based leaning incorporating the position of each node embedded in the WSI is developed and used to classify the prostate and kidney cancer. WSIs are large dimensional, multi-resolution digital slides that store microscopic images of tissues to be used for cancer diagnois.  However, due to their size and complexity, it can be difficult to extract useful information from them using traditional methods such as CNN.

To address this challenge, WSIs are broken into smaller patches. Then, the patches are used to form a graph structure. 
When broken into smaller patches, the position of each patch in the WSI is not encoded in the graph data. This paper takes into account these position embeddings in the training.  This is accomplished with the use of a graph kernel-based technique known as Spline CNN, which incorporates the location of each patch into the graph itself.

An attention-based mechanism is employed for message passing between the nodes in the graph once the position of each patch has been incorporated in the network. Since the attention layer gives varying attention weights to the nodes, the model is able to focus in on specific regions of the image and concentrate on some of them to a greater extent than others. This attention-based strategy is used in combination with graph Grad-CAM to generate heatmaps to identify the regions responsible for the diagnosis. These heatmaps provide visual explanations of the model's prediction and can be compared with the pathologists annotation to better understand and judge the performance of the model. In addition, one can use these heatmaps as initial screening of the pathologist to detect cancerous regions before final cancer grading.

The kappa score, which is a statistical measure of inter-rater agreement, was used to make a comparison between the performance of the position-aware model and that of other approaches. The results of this comparison indicated that the position-embedded method performed better than the other methods when applied to the datasets pertaining to prostate and kidney cancers. 

To address the explianabilty of the approach, this paper tailors the graph Grad-CAM algorithm, which was originally designed for use in other graph learning-based applications, to the proposed self-supervised position-aware algorithm for WSIs. The explainabilty of the model is based on the attention score of the second activation layer of the model. The Grad-CAM approach takes into account each node's attention score before generating a heatmap for that node. The score that is associated with each node is the one that is used for the patch that is associated with that node in the WSI.  All the heatmaps for node is superimposed with the complete WSI to generate the heatmap in the WSI. By visualizing the heatmaps for each node, it is possible to understand which regions of the WSI are most important to the GNN's prediction, and how these regions contribute to the overall prediction made by the model.


\section{Conclusion}
\label{sec:conc}
A self-supervised position-aware attention-based model for cancer grading and classification utilizing WSIs has been developed in this paper. The position embedding of the nodes in the WSI was accomplished by using spline CNN. In order to evaluate the effectiveness of the proposed approach, data sets of WSIs for prostate and kidney cancers were utilized.  The proposed method surpasses established methodologies in cancer diagnosis, such as GCN-based and MIL-based models using kappa score as evaluation metric. In addition to cancer classification and grading, it was shown that the proposed approach is very  explainable and the diagnosis supports the annotated regions in the WSI by the pathologist. 



\bibliographystyle{IEEEbib}
\bibliography{refs}

\begin{thebibliography}{10}

\bibitem{Aeffner2019-vo}
F.~Aeffner, M.~D. Zarella, N.~Buchbinder, M.~M. Bui, M.~R. Goodman, D.~J.
  Hartman, G.~M. Lujan, M.~A Molani, A.~V Parwani, K.~Lillard, O.~C. Turner,
  V.~N.~P. Vemuri, A.G. Yuil-Valdes, and D.~Bowman,
\newblock ``Introduction to digital image analysis in whole-slide imaging: A
  white paper from the digital pathology association,''
\newblock {\em J. Pathol. Inform.}, vol. 10, no. 1, pp. 9, Mar. 2019.

\bibitem{multiinstance}
C.~Mercan, S.~Aksoy, E.~Mercan, L.G. Shapiro, D.L. Weaver, and J.G. Elmore,
\newblock ``Multi-instance multi-label learning for multi-class classification
  of whole slide breast histopathology images,''
\newblock {\em IEEE Transactions on Medical Imaging}, vol. 37, no. 1, pp.
  316--325, 2018.

\bibitem{MIL1}
J.~Melendez, B.~Ginneken, P.~Maduskar, R.~Philipsen, H.~Ayles, and C.~Sanchez,
\newblock ``On combining multiple-instance learning and active learning for
  computer-aided detection of tuberculosis,''
\newblock {\em IEEE Transactions on Medical Imaging}, vol. 35, pp. 1--1, 12
  2015.

\bibitem{mil2}
G.~Quellec, M.~Lamard, M.~Cozic, G.~Coatrieux, and G.~Cazuguel,
\newblock ``Multiple-instance learning for anomaly detection in digital
  mammography,''
\newblock {\em IEEE Transactions on Medical Imaging}, vol. 35, no. 7, pp.
  1604--1614, 2016.

\bibitem{nature-weaklyMIL}
G.~Ampanella, M.G. Hanna, Geneslaw, A.~Miraflor, W.~Krauss, V.~Silva, K.J.
  Busam, E.~Brogi, V.E. Reuter, D.S. Klimstra, and T.J. Fuchs,
\newblock ``Clinical-grade computational pathology using weakly supervised deep
  learning on whole slide images.,''
\newblock {\em Nature Medicine}, 2019.

\bibitem{Lu19}
M.Y. Lu, R.J. Chen, J.~Wang, D.~Dillon, and F.~Mahmood,
\newblock ``Semi-supervised histology classification using deep multiple
  instance learning and contrastive predictive coding,''
\newblock {\em Advances in Neural Information Processing Systems (NeurIPS)
  Workshop in Machine Learning for Health}, 2019.

\bibitem{ma}
M.~Aryal and N.Y. Soltani,
\newblock ``Context-aware graph-based self-supervised learning of whole slide
  images,''
\newblock in {\em IEEE International Conference on Acoustics, Speech and Signal
  Processing (ICASSP)}, 2022.

\bibitem{10.1117/12.2550114}
D.~Anand, S.~Gadiya, and A.~Sethi,
\newblock ``{Histographs: graphs in histopathology},''
\newblock in {\em Medical Imaging 2020: Digital Pathology}, John~E. Tomaszewski
  and Aaron~D. Ward, Eds. International Society for Optics and Photonics, 2020.

\bibitem{chen2021whole}
R.J. Chen, M.Y. Lu, M.~Shaban, C.~Chen, T.Y. Chen, D.F.K Williamson, and
  F.~Mahmood,
\newblock ``Whole slide images are 2d point clouds: Context-aware survival
  prediction using patch-based graph convolutional networks,''
\newblock in {\em Medical Image Computing and Computer Assisted Intervention
  {\textendash} {MICCAI} 2021}, pp. 339--349. Springer International
  Publishing, 2021.

\bibitem{Ding2020FeatureEnhancedGN}
K.~Ding, Q.~Liu, E.H. Lee, M.~Zhou, A.~Lu, and S.~Zhang,
\newblock ``Feature-enhanced graph networks for genetic mutational prediction
  using histopathological images in colon cancer,''
\newblock in {\em MICCAI}, 2020.

\bibitem{NIPS2017_5dd9db5e}
W.~Hamilton, Z.~Ying, and J.~Leskovec,
\newblock ``Inductive representation learning on large graphs,''
\newblock in {\em Advances in Neural Information Processing Systems}, 2017.

\bibitem{kipf2017semi}
T.N. Kipf and M.~Welling,
\newblock ``Semi-supervised classification with graph convolutional networks,''
\newblock in {\em International Conference on Learning Representations (ICLR)},
  2017.

\bibitem{NEURIPS2018_e77dbaf6}
Z.~Ying, J.~You, C.~Morris, X.~Ren, W.~Hamilton, and J.~Leskovec,
\newblock ``Hierarchical graph representation learning with differentiable
  pooling,''
\newblock in {\em Advances in Neural Information Processing Systems}, 2018.

\bibitem{9156339}
Y.~Zhao, F.~Yang, Y.~Fang, H.~Liu, N.~Zhou, J.~Zhang, J.~Sun, S.~Yang,
  B.~Menze, X.~Fan, and J.~Yao,
\newblock ``Predicting lymph node metastasis using histopathological images
  based on multiple instance learning with deep graph convolution,''
\newblock in {\em 2020 IEEE/CVF Conference on Computer Vision and Pattern
  Recognition (CVPR)}, 2020.

\bibitem{801842acbabe476990854c32fdfdbde5}
J.~Wang, R.J. Chen, M.Y. Lu, A.~Baras, and F.~Mahmood,
\newblock ``Weakly supervised prostate tma classification via graph
  convolutional networks,''
\newblock in {\em ISBI 2020 - 2020 IEEE International Symposium on Biomedical
  Imaging}, Apr. 2020.

\bibitem{valentin}
V.~Anklin, P.~Pati, G.~Jaume, B.~Bozorgtabar, A.~Foncubierta-Rodríguez, J.P.
  Thiran, M.~Sibony, M.~Gabrani, and O.~Goksel,
\newblock ``Learning whole-slide segmentation from inexact and incomplete
  labels using tissue graphs,'' 2021.

\bibitem{position}
J.~You, R.~Ying, and J.~Leskovec,
\newblock ``Position-aware graph neural networks,''
\newblock in {\em Proceedings of the 36th International Conference on Machine
  Learning}, Kamalika Chaudhuri and Ruslan Salakhutdinov, Eds., 09--15 Jun
  2019.

\bibitem{splinecnn}
M.~Fey, J.~Lenssen, F.~Weichert, and H.~Muller,
\newblock ``Splinecnn: Fast geometric deep learning with continuous b-spline
  kernels,''
\newblock in {\em 2018 IEEE/CVF Conference on Computer Vision and Pattern
  Recognition (CVPR)}, 2018.

\bibitem{gcam}
R.R. Selvaraju, M.~Cogswell, A.~Das, R.~Vedantam, D.~Parikh, and D.~Batra,
\newblock ``Grad-cam: Visual explanations from deep networks via gradient-based
  localization,''
\newblock in {\em 2017 IEEE International Conference on Computer Vision
  (ICCV)}, 2017.

\bibitem{cam}
C.~Vondrick, A.~Khosla, T.~Malisiewicz, and A.~Torralba,
\newblock ``Hoggles: Visualizing object detection features,''
\newblock in {\em 2013 IEEE International Conference on Computer Vision}, 2013.

\bibitem{cam2}
M.~D. Zeiler and R.~Fergus,
\newblock ``Visualizing and understanding convolutional networks,''
\newblock in {\em Computer Vision -- ECCV 2014}, 2014.

\bibitem{cam3}
K.~Simonyan, A.~Vedaldi, and A.~Zisserman,
\newblock ``Deep inside convolutional networks: Visualising image
  classification models and saliency maps,'' 2013.

\bibitem{pope_explain}
P.~E. Pope, S.~Kolouri, M.~Rostami, C.~E. Martin, and H.~Hoffmann,
\newblock ``Explainability methods for graph convolutional neural networks,''
\newblock in {\em 2019 IEEE/CVF Conference on Computer Vision and Pattern
  Recognition (CVPR)}, 2019.

\bibitem{graphm}
D.~Ahmedt-Aristizabal, M.~A. Armin, S.~Denman, C.~Fookes, and L.~Petersson,
\newblock ``A survey on graph-based deep learning for computational
  histopathology,'' 2021.

\bibitem{graphex}
G.~Jaume, P.~Pati, B.~Bozorgtabar, A.~Foncubierta-Rodríguez, F.~Feroce, A.~M.
  Anniciello, T.~Rau, J.P. Thiran, M.~Gabrani, and O.~Goksel,
\newblock ``Quantifying explainers of graph neural networks in computational
  pathology,'' 2020.

\bibitem{Muja09fastapproximate}
M.~Muja and D.~G. Lowe,
\newblock ``Fast approximate nearest neighbors with automatic algorithm
  configuration,''
\newblock {\em International Conference on Computer Vision Theory and
  Applications}, pp. 331--340, 2009.

\bibitem{chen2020simple}
T.~Chen, S.~Kornblith, M.~Norouzi, and G.~Hinton,
\newblock ``A simple framework for contrastive learning of visual
  representations,''
\newblock {\em Proceedings of the 37th International Conference on Machine
  Learning}, vol. 119, pp. 1597--1607, 13--18 Jul 2020.

\bibitem{fact1}
A.~Ahmed, N.~Shervashidze, S.~Narayanamurthy, V.~Josifovski, and A.~J. Smola,
\newblock ``Distributed large-scale natural graph factorization,''
\newblock in {\em Proceedings of the 22nd International Conference on World
  Wide Web}, New York, NY, USA, 2013.

\bibitem{matfac1}
M.~Belkin and P.~Niyogi,
\newblock ``Laplacian eigenmaps and spectral techniques for embedding and
  clustering,''
\newblock in {\em Proceedings of the 14th International Conference on Neural
  Information Processing Systems: Natural and Synthetic}, Cambridge, MA, USA,
  2001.

\bibitem{deepwalk}
B.~Perozzi, R.~Al-Rfou, and S.~Skiena,
\newblock ``Deepwalk: Online learning of social representations,''
\newblock in {\em Proceedings of the 20th ACM SIGKDD International Conference
  on Knowledge Discovery and Data Mining}, New York, NY, USA, 2014.

\bibitem{node2vec}
A.~Grover and J.~Leskovec,
\newblock ``Node2vec: Scalable feature learning for networks,''
\newblock in {\em Proceedings of the 22nd ACM SIGKDD International Conference
  on Knowledge Discovery and Data Mining}, New York, NY, USA, 2016.

\bibitem{spline}
Z.~Jiang, H.~Rahmani, P.~Angelov, S.~Black, and B.M. Williams,
\newblock ``Graph-context attention networks for size-varied deep graph
  matching,''
\newblock in {\em 2022 IEEE/CVF Conference on Computer Vision and Pattern
  Recognition (CVPR)}, 2022.

\bibitem{spline2}
M.~Rol{\'i}nek, P.~Swoboda, D.~Zietlow, A.~Paulus, V.~Musil, and G.~Martius,
\newblock ``Deep graph matching via blackbox differentiation of combinatorial
  solvers,''
\newblock in {\em Computer Vision -- ECCV 2020}, 2020.

\bibitem{piegl1996nurbs}
L.~Piegl and W.~Tiller,
\newblock {\em The NURBS book},
\newblock 1996.

\bibitem{William}
W.~L. Hamilton,
\newblock ``Graph representation learning,''
\newblock {\em Synthesis Lectures on Artificial Intelligence and Machine
  Learning}, pp. 1--159, 2020.

\bibitem{GAT}
P.~Veli{\v{c}}kovi{\'{c}}, G.~Cucurull, A.~Casanova, A.~Romero, P.~Li{\`{o}},
  and Y.~Bengio,
\newblock ``{Graph Attention Networks},''
\newblock {\em International Conference on Learning Representations}, 2018.

\bibitem{gcam1}
K.~Simonyan, A.~Vedaldi, and A.~Zisserman,
\newblock ``Deep inside convolutional networks: Visualising image
  classification models and saliency maps,'' 2013.

\bibitem{relu}
V.~Nair and G.~E. Hinton,
\newblock ``Rectified linear units improve restricted boltzmann machines,''
\newblock in {\em Proceedings of the 27th International Conference on
  International Conference on Machine Learning}, 2010.

\bibitem{cancer_stat}
R.~L. Siegel, K.~D. Miller, H.~E. Fuchas, and A.~Jemal,
\newblock ``Cancer statistics, 2021,''
\newblock {\em CA: A Cancer Journal for Clinicians}, pp. 7--33, 2021.

\bibitem{TCGA}
O.~Akin, P.~Heller, M.~Jarsosz, B.J. Kirk, and J.~Filippni,
\newblock ``Radiology data from the cancer genome atlas kidney renal clear cell
  carcinoma [tcga-kirc] collection. the cancer imaging archive.,''
\newblock 2016.

\bibitem{Hsieh2017-mm}
J.J. Hsieh, M.~P. Purdue, S.~Signoretti, C.~Swanton, L.~Albiges,
  M.~Schmidinger, D.Y. Heng, J.~Larkin, and V.~Ficarra,
\newblock ``Renal cell carcinoma,''
\newblock {\em Nat. Rev. Dis. Primers}, vol. 3, pp. 17009, Mar. 2017.

\bibitem{kaggle1}
``Artificial intelligence for diagnosis and gleason grading of prostate cancer:
  the panda challenge,''
\newblock {\em NATURE MEDICINE}, vol. 28, pp. 154--163, 2022.

\bibitem{Chen2016-cl}
N.~Chen and Q.~Zhou,
\newblock ``The evolving gleason grading system,''
\newblock {\em Chin. J. Cancer Res.}, vol. 28, no. 1, pp. 58--64, Feb. 2016.

\bibitem{mil_2018}
X.~Wang, Y.~Yan, P.~Tang, X.~Bai, and W.~Liu,
\newblock ``Revisiting multiple instance neural networks,''
\newblock {\em Pattern Recognition}, vol. 74, pp. 15–24, Feb 2018.

\bibitem{mocov3}
X.~Chen, S.~Xie, and K.~He,
\newblock ``An empirical study of training self-supervised vision
  transformers,''
\newblock in {\em 2021 IEEE/CVF International Conference on Computer Vision
  (ICCV)}, 2021.

\bibitem{resnet}
K.~He, X.~Zhang, S.~Ren, and J.~Sun,
\newblock ``Deep residual learning for image recognition,''
\newblock in {\em Proceedings of the IEEE Conference on Computer Vision and
  Pattern Recognition (CVPR)}, June 2016.

\bibitem{adamw}
I.~Loshchilov and F.~Hutter,
\newblock ``Decoupled weight decay regularization,''
\newblock in {\em International Conference on Learning Representations}, 2019.

\bibitem{kappa}
J.~Cohen,
\newblock ``Weighted kappa: Nominal scale agreement with provision for scaled
  disagreement or partial credit.,''
\newblock {\em Psychological Bulletin 70 (4)}, pp. 213--220, 1968.

\end{thebibliography}

\end{document}